\documentclass[12pt,letterpaper]{article}
\usepackage{jheppub}

\title{New Scale Factor Measure}

\author[a,b]{Raphael Bousso,}
\affiliation[a]{Center for Theoretical Physics and Department of Physics,\\
 University of California, Berkeley, CA 94720, U.S.A.}
\affiliation[b]{Lawrence Berkeley National Laboratory, Berkeley, CA 94720,
  U.S.A.}

\abstract{The computation of probabilities in an eternally inflating universe requires a regulator or ``measure''.  The scale factor time measure truncates the universe when a congruence of timelike geodesics has expanded by a fixed volume factor.   This definition breaks down if the generating congruence is contracting---a serious limitation that excludes from consideration gravitationally bound regions such as our own.   Here we propose a closely related regulator which is well-defined in the entire spacetime.  The New Scale Factor Cutoff restricts to events with scale factor below a given value.  Since the scale factor vanishes at caustics and crunches, this cutoff always includes an infinite number of disconnected future regions.  We show that this does not lead to divergences. The resulting measure combines desirable features of the old scale factor cutoff and of the light-cone time cutoff, while eliminating some of the disadvantages of each.}

\begin{document}
\maketitle

\section{Introduction}

%The observed acceleration of the expansion of the universe~\cite{Per98,Rie98} is probably due to a positive cosmological constant~\cite{Bou12}.   Unless the vacuum will soon decay (on a timescale of a few billion years), this implies that the universe as a whole is eternally inflating~\cite{GutWei83}.
  
The observed acceleration of the expansion of the universe~\cite{Per98,Rie98} implies that the universe as a whole is eternally inflating~\cite{GutWei83}.  Absent unnatural tuning, this conclusion is robust: to avoid eternal inflation, our vacuum would have to decay on a timescale of a few billion years.\footnote{Interpretations of the observed acceleration in terms of matter with equation of state parameter different from that of a cosmological constant, or in terms of modifications of general relativity, can be viewed as special cases of this type of tuning, which is strictly in addition to the tuning associated with the smallness of the vacuum energy~\cite{Car00,Bou12}.}

The physics of eternal inflation is straightforward and arises completely within the standard framework of classical gravity coupled to quantum field theory.  Even though the vacuum may only be metastable,%\footnote{Complete stability is excluded by the observed arrow of time~\cite{DysKle02,Bou11,Sus12}.  The same consideration implies that our vacuum was formed by the decay of another metastable vacuum with significantly larger energy.} 
the universe is expanding too rapidly for local decays~\cite{CDL} to lead to a global crunch.  Globally, the universe will become unbounded in size.  Because it is locally at nonzero temperature~\cite{GibHaw77a}, everything that can happen will happen, infinitely many times.  This poses a challenge for the computation of relative probabilities, known as the measure problem.

Eternal inflation is very natural from a theoretical viewpoint.  Slow-roll inflationary models, which are designed to solve the flatness problem and generate density perturbations.  These tasks require mild tuning; but eternal inflation requires no tuning at all.  Moreover, string theory appears to contain a very large number of metastable vacua with either sign of the cosmological constant, a feature that allows it to solve the cosmological constant problem~\cite{BP}.  However, it is worth emphasizing that the measure problem is very robust and has nothing to do with the string landscape.  For eternal inflation to occur, there need only be one single metastable vacuum with positive energy.  Thus the measure problem arises from the most straightforward interpretation of the observed accelerated expansion.

Approaches to the measure problem are inevitably guided by theoretical priors (i.e., by what we perceive as simplicity, elegance, or deep principles).  Yet, much of the recent progress in the subject has come from a hard-nosed investigation of the phenomenology.  A number of proposed measures are in violent conflict with observation and have been eliminated from consideration (see, e.g., ~\cite{FelHal05,BouFre06b,Pag06,BouFre07}).  Among the remaining proposals, some are related by exact dualities; for example, the causal patch cutoff~\cite{Bou06} (with initial conditions in the longest-lived de~Sitter vacuum) is equivalent to the light-cone time cutoff~\cite{Bou09,BouYan09}.  

Here, we focus on the scale factor cutoff~\cite{LinMez93,LinLin94,Lin06,DesGut08a}.  This measure is closely related to the light-cone time cutoff and causal patch.  It has the potential for considerable phenomenological success (see, e.g., Refs.~\cite{FreKle05,Fre08,BouFre08b,DesGut08b,BouHal09,Sal09,BouFre10e}).  However, the definitions provided so far have an unattractive feature: scale factor time is well-defined only in locally expanding regions of spacetime.  In regions undergoing gravitational binding or collapse, such as our own galactic halo, the measure needs to be supplemented with additional rules.  These prescriptions have seemed {\em ad hoc} and entirely unrelated to the definition of scale factor time itself.  Four different possibilities for completions of the scale factor time cutoff were noted in Refs.~\cite{DesGut08a,DesGut08b} alone.\footnote{The most interesting of these proposals, from our viewpoint, is the idea of eliminating the future light-cone of all points on a fixed scale factor time surface.  But then the light-cone time cutoff, which based {\em only\/} on the elimination of future light-cones, would be a simpler, more natural choice.}

In this paper, we propose a New Scale Factor Cutoff, whose definition requires no separate rules for collapsing regions, and we show that it yields a finite probability measure for eternal inflation.

\paragraph{Outline} Given a timelike geodesic congruence, we show in Sec.~\ref{sec-def} that every spacetime point can be uniquely assigned a value of the scale factor, $a$. This relies on a nontrivial result in classical geometry, which guarantees that every point lies on precisely one geodesic with no conjugate points.   It is convenient to work with the scale factor parameter $\eta= \log a$. (We refer to $\eta$ as a parameter since it is not, generally, a time function.) We define the New Scale Factor Cutoff as a restriction to spacetime points with scale factor parameter below a given value of $\eta$.  As usual, relative probabilities are computed in this finite region before taking the limit $\eta\to\infty$.

\begin{figure}[tbp]\centering
\includegraphics[width=.9 \textwidth]{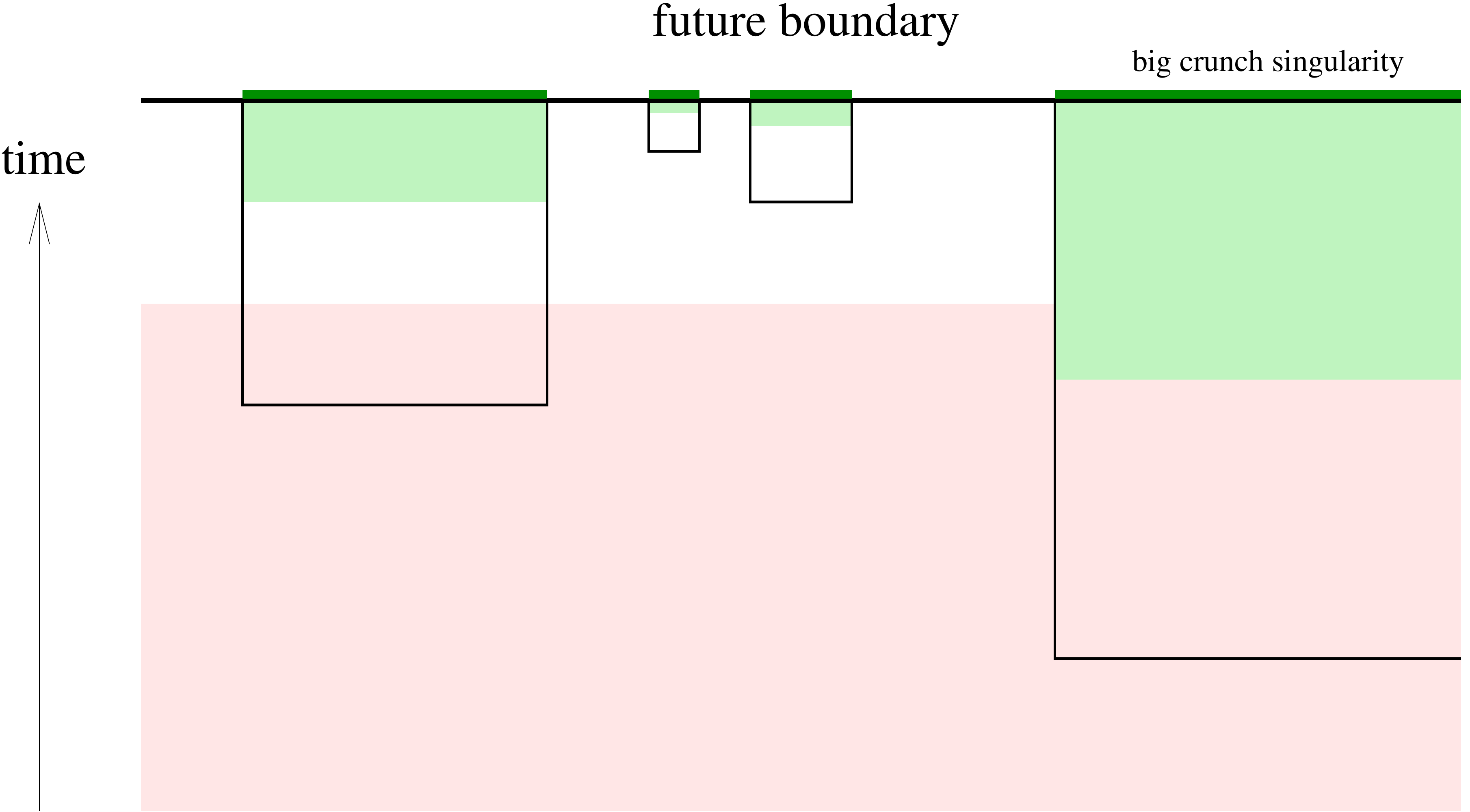}
\caption{Schematic depiction of the New Scale Factor Cutoff in a simple model with one de~Sitter vacuum and one vacuum with a big crunch. (This model is considered in detail in Sec.~5.)  Square bubbles of the crunching vacuum are shown; the crunch is at the future boundary (thick green intervals).  Shaded regions correspond to the cutoff region $M(\eta)$.  The expanding portion of $M(\eta)$ is connected (pink).  The contracting portion of $M(\eta)$ (green) consists of an infinite number of disconnected regions.  The later a bubble nucleates, the thinner the cutoff region near the caustics.   As a result, the total contribution from the collapsing regions is convergent.}
\label{fig-square}
\end{figure}
The scale factor parameter decreases in collapsing regions, and it diverges to $-\infty$ at caustics and singularities.  Formally, an infinite number of disconnected collapsing regions must therefore be included at any finite cutoff (Fig.~\ref{fig-square}).  In the remainder of the paper, we show that neither of these unconventional features leads to any problems.  An important point is that a useful geometric cutoff should select a finite portion of the infinite spacetime, but {\em this portion need not be connected, nor does it need to be bounded by an ``instant of time''} (a spacelike or null hypersurface).\footnote{Nothing is lost by abandoning the latter criterion, since it is not generally satisfied even by the old definition.  Hypersurfaces of constant (old) scale factor time can contain timelike portions.}  Abandoning connectedness is a key feature that distinguishes the New Scale Factor Cutoff from the old scale factor measure, allowing it reach all points in the spacetime without having to invent separate {\em ad hoc\/} rules for collapsing regions.

In Sec.~\ref{sec-rate}, we derive the rate equation for the distribution of de~Sitter vacua as a function of the scale factor parameter $\eta$.  We use standard approximations, in which all de~Sitter vacua are treated as empty and all domain walls as comoving. In this approximation, the scale factor parameter is monotonically increasing.  Thus, the rate equation is identical to that of old scale factor time.  At late times, the evolution is dominated by an attractor regime controlled by the leading eigenvector of the rate equation.

In Sec.~\ref{sec-ni}, we compute the number $N_I(\eta)$ of events of arbitrary type $I$, as a function of the cutoff parameter $\eta$, in the cutoff region.  At this stage nonempty regions, and in particular collapsing regions, must be treated in full.  We show that only a finite number of collapsing regions contribute to $N_I(\eta)$ at any value of the cutoff, and that an attractor regime is reached as for other global cutoffs.  Thus, the New Scale Factor Cutoff yields well-defined probability amplitudes:
\begin{equation}
\frac{P_I}{P_J}=\lim_{\eta\to\infty} \frac{N_I(\eta)}{N_J(\eta)}~.
\end{equation}

In Sec.~\ref{sec-example}, we compute probabilities in a concrete toy model of an eternally inflating universe that contains collapsing regions.  This example illustrates explicitly that the novel features of New Scale Factor Cutoff do not lead to pathologies.

\paragraph{Discussion} We leave the detailed investigation of the phenomenology of the New Scale Factor Cutoff to future work.  One expects, however, that the New Scale Factor Cutoff will share some of the desirable features of the (old) scale factor time measure~\cite{BouFre08b,DesGut08a,DesGut08b,DesSal09} and of its local dual, the fat geodesic~\cite{BouFre08b,LarNom11,Sal12}).  These features include the absence  (subject to plausible assumptions on the structure of the vacuum landscape) of catastrophies such as the youngness paradox and the dominance of Boltzmann brains (see, e.g., Refs.~\cite{Teg05,BouFre06b,SchVil06,BouFre07,Fre11,Sal11}). Compared to the causal patch measure, which shares the above features, the New Scale Factor Cutoff is likely to suppress vacua with $\Lambda<0$ more strongly~\cite{Sal09}, and it may not diverge in ``hat regions'' with $\Lambda=0$.

The New Scale Factor Cutoff is well-defined in collapsing regions, but its evaluation in realistic models could be challenging.  It requires detailed tracking of a geodesic congruence and identifying its complicated caustic surfaces. The fact that the causal patch depends only on the endpoint of the geodesic (and might not require a congruence at all~\cite{BouFre10b}) remains a significant formal and practical advantage.

A particular (and, so far, unique) phenomenological advantage of the causal patch measure~\cite{BouHar07,BouHar10,BouFre10d} is its robust prediction of the observed value of the cosmological constant (at least over positive values of $\Lambda$; see~\cite{Sal09}), eliminating the need to impose  anthropic conditions such as the presence of galaxies~\cite{Wei87,MarSha97,DesGut08a} and sufficient metallicity~\cite{LarNom11}.  The causal patch explains the coincidence of vacuum and matter energy directly: the value of $\Lambda$ is determined by the the timescale when observers exist.  It will be interesting to study whether the New Scale Factor Cutoff is able to reproduce this success.\footnote{I am grateful to B.~Freivogel for discussions of this question.}

\section{New Scale Factor Cutoff}
\label{sec-def}
In the eternally inflating spacetime, consider a smooth ($C^\infty$) spacelike hypersurface $\Sigma_0$.  Every point $x_0\in \Sigma_0$ is the starting point of a future-directed timelike geodesic orthogonal to $\Sigma_0$.  We assume that the future of $\Sigma_0$ is eternally inflating, or more precisely, that there exists a set (of measure zero, but nonempty) of orthogonal geodesics that remain forever in a de Sitter phase.  This will usually be the case if $\Sigma_0$ contains much more than one horizon volume of metastable de~Sitter vacua~\cite{GutWei83}. 

We will distinguish between a family and a congruence of geodesics.  A congruence in the spacetime subset $M$ is a family of geodesics such that for every point $p\in M$, there is precisely one geodesic through $p$.  The family of geodesics orthogonal to $\Sigma_0$ will be a congruence initially~\cite{Wald} but will cease to be a congruence after geodesics intersect.  Intersections can be local or nonlocal.  A local intersection is where infinitesimally neighboring geodesics meet, and the local expansion $\theta$ diverges to $-\infty$.  This is called a caustic, or focal point, or conjugate point.  A nonlocal intersection is where nonneighboring geodesics meet, i.e., geodesics whose origins on $\Sigma_0$ are more than infinitesimally apart.  Because of the accelerated expansion driving eternal inflation, geodesics emanating from distinct points on $\Sigma_0$ will either intersect nonlocally very close to $\Sigma_0$\footnote{One expects that this can be avoided by a judicious choice of $\Sigma_0$.} or fall out of causal contact.  Thus, nonlocal intersections will not play a role at late times, i.e., in the limit $\eta\to\infty$ in which probabilities will be defined.

This distinction between a family and a congruence can be important.  For example, the light-cone time cutoff is defined using a {\em family\/} of geodesics orthogonal to $\Sigma_0$, all of which are maximally extended, whether or not they intersect other geodesics~\cite{Bou09}.  The New Scale Factor Cutoff, on the other hand, will crucially require the use of a congruence.  

We can easily obtain a congruence from the geodesics orthogonal to $\Sigma_0$, by declaring that {\em every geodesic will be terminated at its first conjugate point.}  This ensures (absent nonlocal intersections) that no point is contained on more than one geodesic.  To each point $p$ in the congruence, we can thus uniquely assign a scale factor parameter $\eta$, defined in terms of an integral along the unique geodesic from $\Sigma_0$ to $p$:
\begin{equation}
\eta\equiv \int \frac{\theta(t)}{3} dt~.
\label{eq-sft}
\end{equation}
Here, $t$ is the proper time along the geodesic;
\begin{equation}
\theta\equiv \frac{d}{dt}\log\frac{dV}{dV_0}
\label{eq-exp}
\end{equation}
is the expansion of the congruence; $dV$ is the volume element at the proper time $t$ along a geodesic spanned by infinitesimally neighboring geodesics in the congruence; and $dV_0$ is the volume element spanned by the same neighbors at $t=0$.  In terms of the unit tangent vector field (the four-velocity) of the geodesic congruence, $\xi=\partial_t$, the expansion can be computed as~\cite{Wald}
\begin{equation}
\theta=\nabla_a\xi^a~.
\end{equation}

We have defined $\eta$ so that it vanishes at $\Sigma_0$.  At later times along a given geodesic, $\eta$ will increase while the congruence is expanding ($\theta>0$), and $\eta$ will decrease during contracting phases.  In particular, $\eta$ is not of definite sign, and it diverges to $-\infty$ as a conjugate point is approached.  For simplicity, we will assume the generic case in which there is at most one contracting phase, which leads to a caustic.  Because the strong energy condition is not satisfied in regions with positive cosmological constant, the congruence may in principle expand and contract multiple times.  This case presents no particular difficulties and could easily be included.  However, it is not known to arise at late times in realistic models, and it would make our discussion and notation more cumbersome.

Some geodesics will never reach a caustic (for example, if they enter a terminal vacuum with vanishing cosmological constant). Other geodesic may avoid caustics until they encounter a singularity, such as a black hole singularity or a big crunch in a vacuum region with negative cosmological constant.  Such geodesics are maximally extended.  But in general, caustics do occur in regular spacetime regions, particularly in gravitationally bound structures such as our galactic halo.  Geodesics encountering caustics will be terminated early; they are extendible.  

This raises an important question: will the congruence have ``gaps''---will there be points that would have been reached by the fully extended geodesics, but which do not lie on any geodesic once we impose the rule that geodesics are terminated at conjugate points?

In fact, the congruence has no gaps.  This is guaranteed by a result in classical geometry.  Without reproducing the proof, we quote the Corollary to Lemma 6.7.3 in the textbook by Hawking and Ellis~\cite{HawEll}:
\begin{quote}
If $\cal S$ is a $C^2$ partial Cauchy surface, then to each point $q\in D^+({\cal S})$ there is a future-directed timelike geodesic curve orthogonal to $\cal S$ [...] which does not contain any point conjugate to $\cal S$ between $\cal S$ and $q$.
\end{quote}
A partial Cauchy surface is spacelike hypersurface which no causal curve intersects more than once.  Therefore, we may apply this result to ${\cal S}=\Sigma_0$.  Strictly, it guarantees the absence of gaps only in the future domain of dependence $D^+$ of $\Sigma_0$ (the set of points $p$ such that every inextendible past-directed causal curve from $p$ intersects $\Sigma_0$).  Some of the outermost geodesics in the congruence do not remain within this set.  However, one would expect a more general result to hold.  In any case, we have already assumed that $\Sigma_0$ contains more than one de~Sitter horizon volume.  Then $D^+(\Sigma_0)$ alone contains an eternally inflating universe, and in the limit as $\eta\to\infty$, effects from the edge of the congruence will become arbitrarily suppressed.   Therefore, the quoted Corollary is sufficient.

We conclude that the scale factor parameter is well-defined in all spacetime regions, expanding or collapsing.  {\em The New Scale Factor Cutoff is the restriction of the eternally inflating spacetime to the set of points $M(\eta)$ with scale factor parameter less than $\eta$.}

We will now specify the associated probability measure.  For this purpose, we make a number of definitions which have become standard in the application of geometric cutoffs to the measure problem.  Let $I$ and $J$ be possible outcomes of some measurement.\footnote{For example, each could represent some range of values for the Higgs mass, or for the cosmological constant.  We are free to condition on as many things as we like, such as other known parameters, the precise nature of the experiment, or the environment in which it is carried out.  Such conditions are part of the definition of $I$ and $J$.   As always, the less we choose to condition on, the more powerful the prediction or postdiction.---As a side remark, we note the freedom to choose the question we wish to ask is unrelated to the language in which we cast the question, whether it is classical, or quantum-mechanical.  (It is usually assumed implicitly that the number of events of type $I$ is more fundamentally an expectation value computed in an appropriate quantum state; see, e.g., Ref.~\cite{BouSus11} for a discussion of suitably collapsed states arise in the causal patch measure.)  And in neither language will any amount of conditioning alleviate the measure problem: in an eternally inflating universe, the past light-cone of any event has finite maximal area, and thus contains a finite amount of information~\cite{CEB1,BouFre10a}.  Because of the unbounded expansion, there exist infinitely many causally disconnected points whose past light-cones contain exactly the same quantum state.  Therefore, no amount of conditioning can localize an experiment in the multiverse.  Unless the outcome of the experiment is trivially determined by the specified conditions (e.g., if the Higgs mass is measured but also conditioned on), every possible outcome will occur in infinitely many locations.} Let $N_I(\eta)$ be the number of times outcome $I$ occurs in the spacetime region $M(\eta)$.   The relative probability of the outcomes $I$ and $J$ is defined by
\begin{equation}
\frac{P_I}{P_J}=\lim_{\eta\to\infty} \frac{N_I(\eta)}{N_J(\eta)}~.
\label{eq-nsf}
\end{equation}
Normalized probabilities can be obtained as usual by computing relative probabilities for a complete set of alternative outcomes.

Because $\eta$ is not monotonic along the congruence, $M(\eta)$ may consist of multiple disconnected spacetime regions.  In fact, the number of disconnected components will be infinite, if the eternally inflating spacetime includes regions where the congruence contracts.  Examples of such regions include pocket universes with negative cosmological constant, and regions where gravitational binding occurs, such as the region we live in.

In the remainder of this paper we will examine the properties of the New Scale Factor Cutoff in detail. For any value of $\eta$, we will find that effectively, only a finite number of the disconnected of components of $M(\eta)$ contribute to $N_I(\eta)$.   More precisely, the total contribution from the infinite number of late bubbles converges to a finite result. Thus, the probability measure, Eq.~(\ref{eq-nsf}), is well-defined and can in principle be computed from parameters of the theory.

\section{Rate Equation}
\label{sec-rate}

Our task is to compute for the number of events of type $I$, $N_I(\eta)$, that occur in the cutoff region $M(\eta)$ defined in the previous section.  It is convenient to do this in two steps.  In this section, we perform the first step, which focusses on the dynamics of eternal inflation.  We restrict attention to regions occupied by de~Sitter vacua (metastable vacua with positive cosmological constant).  Moreover, we approximate such vacua as entirely devoid of matter.  Further approximations will be spelled out below. The resulting spacetime can be thought of as a scaffold to which, in Sec.~\ref{sec-ni}, we will add matter and terminal vacua (regions with nonpositive cosmological constant).

This division is convenient in the evaluation of global cutoffs~\cite{GarSch05,DesGut08a,BouFre07,BouFre08b,BouYan09}, because the timescales for vacuum decay are much larger than the matter or radiation eras immediately following decay.  It offers an additional advantage in the case of the New Scale Factor Cutoff: the scaffold we will build contains no contracting regions.  Therefore, for the purposes of this section, $M(\eta)$ will be a single connected region.  It will be bounded in the past by $\Sigma_0$, and in the future by a hypersurface $\Sigma_\eta$.  Our task will be to compute the volume $V_\alpha(\eta)$ occupied by the de~Sitter vacuum $\alpha$.\footnote{We will use Greek indices to label de~Sitter vacua ($\Lambda>0$), indices $m,n,\ldots$ to label terminal vacua ($\Lambda\leq 0$), and $i,j$ for arbitrary vacua.}

In regions occupied by vacuum $\alpha$, the local metric rapidly approaches the form
\begin{equation}
ds^2=-dt^2+e^{2t/t_{\Lambda,\alpha}} d {\mathbf x}^2~,
\label{eq-ds}
\end{equation}
where $t$ is proper time, and
\begin{equation}
t_{\Lambda,\alpha} \equiv \sqrt{\frac{3}{\Lambda_\alpha}}
\end{equation}
is the timescale associated with the cosmological constant.  At the same rate, exponentially in $t/t_{\Lambda,\alpha}$, geodesics of any congruence become comoving with the arbitrary coordinate system of Eq.~(\ref{eq-ds}).  Therefore, the local expansion is $\theta= t_{\Lambda,\alpha}^{-1}$, and the relation between proper time and scale factor time is
\begin{equation}
d\eta=\frac{dt}{t_{\Lambda,\alpha}}~,
\label{eq-detadt}
\end{equation}
and the metric can be written locally as
\begin{equation}
ds^2=-dt^2+e^{2\eta} d {\mathbf x}^2~,
\label{eq-dseta}
\end{equation}
up to shifts in $\eta$ that can be absorbed into rescalings of the Euclidean spatial coordinates ${\mathbf x}$.  Since slices of constant scale factor time are spatially flat on the horizon scale, a flat horizon patch occupies a physical volume
\begin{equation}
v_\alpha=\frac{4\pi}{3}t_{\Lambda,\alpha} ^3~.
\end{equation}

The rate equation for the volume distribution of de~Sitter vacua is
\begin{equation}
\frac{dV_\alpha}{d\eta} = (3-\kappa_\alpha) V_\alpha + \sum_\beta \kappa_{\alpha\beta} V_\beta~,
\label{eq-vrate}
\end{equation}
where $\kappa_{i\beta}=v_\alpha t_{\Lambda,\beta} \Gamma_{i\beta}$ is the dimensionless decay rate from $\beta$ to $i$.  That is, $\Gamma_{i\beta}$ is the rate at which $i$-bubbles are produced inside the $\beta$-vacuum, per unit four-volume; and $\kappa_{i\beta}$ is the decay rate per unit horizon volume and unit de~Sitter time scale.  By $\kappa_\alpha\equiv \sum_i \kappa_{i\alpha}$ we denote the total dimensionless decay rate of vacuum $\alpha$.  We will now explain the origin of each term on the right-hand side.

The first term, $3 V_\alpha$, follows directly from the exponential volume growth of de~Sitter space, Eq.~(\ref{eq-dseta}).  This would be the only term were it not for dynamical transitions between vacua.

The second term, $-\kappa_\alpha V_\alpha$ is an effective term that takes into account the decay of vacuum $\alpha$ into other vacua.  Decays of this type proceed by the formation of a bubble of the new vacuum~\cite{CDL}.  Typically, the spherical domain wall separating the vacua will be small initially, compared to the size of the event horizon of the parent vacuum.   The domain wall will then expand at a fixed acceleration, asymptotically approaching the future light-cone of the nucleation event.  A detailed treatment of this dynamics would enormously complicate the rate equation, but fortunately an excellent approximation is available.  Because of de~Sitter event horizons, only a portion the of parent vacuum is ever destroyed by the bubble.  This portion is the causal future of the nucleation event, but at late times it agrees with the {\em comoving\/} future of a single horizon volume centered on the nucleation point, at the nucleation time.   Moreover, the bubble reaches its asymptotic comoving size very quickly, exponentially in light-cone time. Thus only a very small error, of order $\kappa_\alpha$, is introduced if we remove this comoving future, rather than the causal future, from the parent vacuum.  That is, for every decay event in vacuum $\alpha$, we instantly reduce the volume $V_\alpha$ by one horizon volume, $v_\alpha$, in the rate equation.  This is called the {\em square bubble approximation}. The infinitesimal number of decay events is $\sum_i \Gamma_{i\alpha} V_\alpha dt=\frac{\kappa_\alpha}{v_\alpha} V_\alpha  d\eta$.  Multiplication by $v_\alpha$ yields the infinitesimal volume loss.

The third term, $\sum_\beta \kappa_{\alpha\beta} V_\beta$, captures the production of bubbles of vacuum $\alpha$ by the decay of other vacua.  In the square bubble approximation, the prefactor of this term must be fixed by demanding continuity of the scale factor parameter across nucleation events.  By the arguments of the previous paragraph, the $\beta$-volume that is instantaneously lost per $\alpha$-nucleation is $v_\beta$, and it must be replaced by an equal volume of $\alpha$-vacuum.  The infinitesimal amount of $\alpha$-volume added by decays in the time $d\eta$ is obtained by computing the number of such decay events $dN_{\alpha\beta}=\Gamma_{\alpha\beta} V_\beta dt$, multiplying by $v_\beta$, and summing over all $\beta$.

The rate equation (\ref{eq-vrate}) has the solution~\cite{GarSch05}
\begin{equation}
V_\alpha(\eta)=\bar V_\alpha e^{\gamma\eta}+O(e^{\varphi\eta})~.
\label{eq-vattract}
\end{equation}
where $\varphi<\gamma< 3$.  (The case $\gamma = 3$ arises if and only if the landscape contains no terminal vacua, i.e., vacua with nonpositive cosmological constant, and will not be considered in this paper.)  Here, $\gamma\equiv 3-q$ is the largest eigenvalue of the matrix $M_{\alpha\beta}$ defined by rewriting Eq.~(\ref{eq-vrate}) as
\begin{equation}
\frac{dV_\alpha}{d\eta}=\sum_\beta M_{\alpha\beta} V_\beta~;
\end{equation}
and $\bar V_\alpha$ is the corresponding eigenvector.  The (square) transition matrix is given by the expression
\begin{equation}
M_{\alpha\beta}=\kappa_{\alpha\beta}-\delta_{\alpha\beta} \kappa_\alpha~.
\end{equation}

The terms of order $e^{\varphi t}$ are subleading and become negligible in the limit as $\eta \to \infty$.  To a very good approximation (better than $q\ll 1$)~\cite{SchVil06}, the eigenvector is dominated by the longest-lived metastable de~Sitter vacuum in the theory, which will be denoted by $*$:
\begin{equation}
\bar V_\alpha\approx \delta_{\alpha*}~,
\end{equation}
and
\begin{equation}
q\approx \kappa_*
\end{equation}
is its total dimensionless decay rate.

\section{Attractor Solution}
\label{sec-ni}

In this section, we compute the number $N_I(\eta)$ of events of type $I$ in the New Scale Factor Cutoff region $M(\eta)$.  At this stage, we will need to confront the infinite number of disconnected collapsing regions that are contained in $M(\eta)$ at any finite value of $\eta$.  Before addressing this novel issue in Sec.~\ref{sec-collapse}, however, it is instructive to derive $N(\eta)$ in the idealized case where the congruence is everywhere expanding.  

\subsection{Attractor Solution Without Collapsing Regions}
\label{sec-nocollapse}

We assume that the events unfolding in a new bubble of vacuum $i$ depend only on $i$, but not on the time of nucleation.  This is true as long as the parent vacuum is long-lived, so that most decays occur in empty de~Sitter space.  For notational convenience, we will also assume that evolution inside a new bubble is independent of the parent vacuum; however, this dependence could easily be included in the analysis.

Then the number of events of type $I$ inside a bubble of type $i$, $dN_I/dN_i$ will depend only on the scale factor time since bubble nucleation, $\zeta\equiv \eta -\eta_{\rm nuc}$.  Therefore, we can write
\begin{equation}
N_I(\eta)=\frac{\kappa_{I*}}{v_*} V_*(\eta)+\sum_{i\neq *}\int_0^{\eta}\left(\frac{dN_I}{dN_i}\right)_{\eta-\eta_{\rm nuc}} \left(\frac{dN_i}{d \eta }\right)_{\eta_{\rm nuc}} d \eta_{\rm nuc} ~,
\label{eq-nilct1}
\end{equation}
Because the dominant vacuum $*$ plays a role analogous to an equilibrium configuration, it is convenient to separate it out from the sum, and to define $\kappa_{I*}$ as the dimensionless rate at which events of type $I$ are produced in $*$ regions.  The rate at which vacua of type $i$ are nucleated is
\begin{equation}
\frac{dN_i}{d \eta} = \sum_\beta \frac{\kappa_{i\beta}}{v_\beta} V_\beta(\eta)~.
\label{eq-sfrate}
\end{equation}
Note that the natural quantity appearing in the above equation is not the volume but the number of horizon patches, $V_\alpha(\eta)/v_\alpha$.  Therefore, it will be  convenient to define 
\begin{equation}
\bar n_\alpha\equiv \frac{\bar V_\alpha}{v_\alpha}~.
\label{eq-barni1}
\end{equation}
By Eq.~(\ref{eq-vattract}), the number of horizon patches of type $i$ at scale factor time $\eta$ obeys
\begin{equation}
n_\alpha(\eta)=\bar n_\alpha e^{\gamma\eta}+O(e^{\varphi\eta})~.
\label{eq-nibarni}
\end{equation}

By changing the integration variable to $\zeta$ in Eq.~(\ref{eq-nilct1}), and using Eq.~(\ref{eq-nibarni}), one finds that 
\begin{equation}
N_I(\eta)=\left(\kappa_{I*} \bar n_*+\sum_{i\neq *}\sum_\beta N_{Ii} \kappa_{i\beta} \bar n_\beta\right) e^{\gamma \eta}+O(e^{\varphi \eta})~,
\label{eq-NIsolution}
\end{equation}
where
\begin{equation}
N_{Ii}\equiv \int_0^\infty d\zeta e^{-\gamma \zeta} \left( \frac{dN_I}{dN_i}\right)_{\zeta}
\label{eq-nii}
\end{equation}
depends only on $I$ and $i$.  The above integral runs over the interior of one $i$-bubble, excluding regions where $i$ has decayed into some other vacuum.  Naively, the integral should range from $0$ to $\eta$.  But the global measure requires us to take the limit $\eta\to\infty$ in any case, and it can be done at this step separately without introducing divergences.  Since $*$ does not appear in the sum in Eq.~(\ref{eq-NIsolution}), and all other vacua decay faster than $*$, the interior of the $i$-bubble in Eq.~(\ref{eq-nii}) grows more slowly than $e^{\gamma \eta}$.  Therefore, the integral converges, and we may write
\begin{equation}
N_I(\eta )=\bar N_I e^{\gamma \eta}+O(e^{\varphi \eta})~,
\label{eq-sfattract}
\end{equation}
where
\begin{equation}
\bar N_I\equiv \kappa_{I*} \bar n_*+\sum_{i\neq *}\sum_\beta N_{Ii} \kappa_{i\beta} \bar n_\beta~.
\label{eq-barni}
\end{equation}

The subleading terms become negligible in the limit as $\eta\to\infty$.  By Eq.~(\ref{eq-nsf}), relative probabilities are given by
\begin{equation}
\frac{P_I}{P_J}=\frac{\bar N_I}{\bar N_J}~,
\label{eq-nsftwo}
\end{equation}
The unnormalized probabilities $\bar N_I$ and $\bar N_J$ are given by Eq.~(\ref{eq-barni}) and can be computed by standard methods, given the theory.

\subsection{Including Collapsing Regions}
\label{sec-collapse}

When collapsing regions are included, the main new feature is that scale factor time will not increase monotonically along the geodesics.  For simplicity we will focus on the simplest (and generic) case where the expansion changes sign only once, at a maximum of the scale factor.  (Since the strong energy condition is not assumed, the general case involves local bounces but this can easily be included.)   

Let us follow an arbitrary geodesic orthogonal to the initial surface.  We set $\eta=0$ at the starting point $x_0\in\Sigma_0$.  While the expansion $\theta$ is positive, $\eta$ increases.  It reaches a maximum, $\eta_{\rm max}$, where $\theta=0$, and then begins to decrease.  By Eq.~(\ref{eq-exp}), the scale factor parameter diverges to $-\infty$ at the first caustic.  Therefore, a portion of {\em every\/} collapsing region will be included at any finite scale factor cutoff $\eta$.  But if there are any collapsing regions at all, then there will be infinitely many such regions in the future of $\Sigma_0$.   This would seem to make $N_I(\eta)$ poorly defined and possibly divergent.  

However, let us terminate each geodesic slightly early, say one Planck time {\em before\/} a caustic or singularity is reached.\footnote{In fact, the example in Sec.~\ref{sec-example} shows that it is not necessary to impose this additional cutoff.  The total contribution from the infinite number of collapsing regions converges even without a cutoff near caustics and singlarities.  We introduce it here because it is physically reasonable and it simplifies the argument.}   At the semiclassical level, this is physically indistinguishable from the definition of the New Scale Factor Cutoff given earlier, but mathematically it greatly simplifies the analysis.  It renders finite the amount by which $\eta$ decreases during the collapse:
\begin{equation}
\Delta\eta_-\equiv \eta_{\rm max}-\eta_\dagger<\infty~.
\end{equation}
Here, $\eta_\dagger$ is the value of the scale factor parameter at the regulated endpoint of the geodesic.

We now come to the crucial point: by Eq.~(\ref{eq-sft}), $\Delta\eta_-$ depends only on local physics in the collapsing region, but not on $\eta_{\rm max}$.  Similarly, the difference between the turnaround time and the nucleation time of the bubble containing the collapsing region,
\begin{equation}
\Delta\eta_+\equiv \eta_{\rm max}-\eta_{\rm nuc}~,
\end{equation}
depends only on the properties of the pocket universe.  We conclude that the scale factor parameter increases and decreases by the same finite amounts during the expansion and collapse phases in a particular pocket universe, no matter when that that bubble universe is nucleated.  This implies that only a finite number of bubbles need be considered at any value of the cutoff.

To see this explicitly, let us define
\begin{equation}
\Delta\eta\equiv \Delta\eta_- -\Delta\eta_+~,
\end{equation}
and let $\Delta\eta_{\rm sup}$ be the smallest upper bound on $\Delta\eta$, among all geodesics in the congruence, or (for later convenience) at least zero:
\begin{equation}
\Delta\eta_{\rm sup}\equiv\min\{0,\sup_{x_0} \Delta\eta\}~.
\end{equation}
On physical ground one expects that $\Delta\eta_{\rm sup}<\infty$.  

Let $Q$ be an event in an empty de~Sitter region, with scale factor time $\eta_Q$, and consider the short fat geodesic through $Q$.  By the definition of $\Delta\eta_{\rm sup}$, the values of $\eta$ along this geodesic in the future of $Q$ are guaranteed to satisfy
\begin{equation}
\eta>\eta_Q - \Delta\eta_{\rm sup}
\end{equation}
(At this point we make use of the nonnegativity of $\Delta\eta_{\rm sup}$.)
Therefore, in the analogue of Eq.~(\ref{eq-nilct1}) for scale factor time,
\begin{equation}
N_I(\eta)=\kappa_{I*} n_*+\sum_{i\neq *}\int_0^{\eta+\Delta\eta_{\rm sup}}
\left(\frac{dN_I}{dN_i}\right)_{\eta-\eta_{\rm nuc}} 
\left(\frac{dN_i}{d \eta }\right)_{\eta_{\rm nuc}} d \eta_{\rm nuc} ~.
\label{eq-nisft1}
\end{equation}
Through $dN_I/dN_i$, this integral includes collapsing regions in the future of the hypersurface of old scale factor time $\eta$, including, if $\Delta\eta_{\rm sup}>0$, collapsing regions in some bubbles that have not yet formed by this time. But we need only integrate over a {\em finite\/} number of bubbles, namely the bubbles that form before the time $\eta+\Delta\eta_{\rm sup}$.  

The cumulative number of events of type $I$ inside a bubble of type $i$, $dN_I/dN_i$, will still depend only on the difference $\zeta=\eta-\eta_{\rm nuc}$ between the New Scale Factor Cutoff and the time of bubble nucleation.  But instead of requiring $\zeta>0$, we now relax this condition to $\zeta>-\Delta\eta_{\rm sup}$.  Correspondingly, the analogue of Eq.~(\ref{eq-nii}) becomes
\begin{equation}
N_{Ii}\equiv \int_{-\Delta\eta_{\rm sup}}^\infty d\zeta e^{-\gamma \zeta} \left( \frac{dN_I}{dN_i}\right)_{\zeta}
\label{eq-nii2}
\end{equation}

As in the previous subsection, this integral converges because all vacua decay faster than the dominant vacuum, and one obtains the same attractor behavior, Eq.~(\ref{eq-sfattract}).  Relative probabilities are again given by
\begin{equation}
\frac{P_I}{P_J}=\frac{\bar N_I}{\bar N_J}~,
\label{eq-nsfthree}
\end{equation}
where
\begin{equation}
\bar N_I\equiv \kappa_{I*} \bar n_*+\sum_{i\neq *}\sum_\beta N_{Ii} \kappa_{i\beta} \bar n_\beta~.
\label{eq-barnithree}
\end{equation}

\section{Example}
\label{sec-example}

In this section, we compute probabilities in a simple example.  Its main purpose is to demonstrate explicitly that the New Scale Factor Cutoff is finite, even though it includes an infinite number of collapsing regions at finite cutoff.   

Consider a toy model with just one metastable de~Sitter vacuum (which is automatically the dominant vacuum), and one terminal vacuum containing a collapsing phase (as would be the case if it had negative cosmological constant).  The metric in the de~Sitter region is
\begin{equation}
ds^2=-dt^2+H^{-2} e^{2Ht}d {\mathbf x} ^2= H^{-2} (-d\eta^2+e^{2\eta} d {\mathbf x}^2)~.
\end{equation}
We take $\Sigma_0$ to be a large but finite portion of the hypersurface $t=\eta=0$, for example a sphere of radius $|{\mathbf x}|<10$.

We treat the terminal pocket universes as true square bubbles.  That is, when a bubble of terminal vacuum nucleates, we take it to instantaneously occupy one horizon volume of the de~Sitter parent vacuum.  This is depicted in Fig.~\ref{fig-square}.  In the more general discussion in the previous sections, the square bubble approximation was applied only in the derivation of the rate equation (Sec.~\ref{sec-rate}).  The purpose of the present example is not to work out a realistic model to great precision, but to illustrate how the New Scale Factor Cutoff operates in the presence of collapsing reasons.  Square bubbles eliminate complications that are unrelated to this central issue.  

Similar considerations will guide our choice of scale factor. The metric inside a bubble of terminal vacuum,
\begin{equation}
ds^2=-d\tau^2+a^2(\tau) d {\mathbf x}^2~,
\end{equation}
will be a mock-up of a flat universe that first expands and then recollapses.  The scale factor $a$ breaks up into factors before and after nucleation:
\begin{equation}
a(\tau)=e^{\eta(\tau)}=e^{\eta_{\rm nuc}+\zeta(\tau)}
\end{equation}
For the latter, we choose linear expansion and contraction (after an initial transitory phase of duration $\epsilon$):
\begin{equation}
e^{\zeta(\tau)}=\left\{\begin{array}{ll}
e^{H\tau}~, & 0<\tau<\epsilon \\
c_+\tau~, & \epsilon<\tau<\tau_{\rm turn} \\
c_-\tilde\tau~, & 0<\tilde\tau<\tilde\tau_{\rm turn}~.
\end{array}\right.
\label{eq-zeta}
\end{equation}
Here, $\tau$ is the proper time since nucleation; $\tau_{\rm turn}$ is the time of maximum expansion; and 
\begin{equation}
\tilde\tau\equiv \tau_{\rm crunch}-\tau
\end{equation}
is the time remaining before the big crunch, $a\to 0$.  The crunch thus occurs at the time $\tilde\tau=0$ or $\tau=\tau_{\rm crunch}=\tau_{\rm turn}+\tilde\tau_{\rm turn}$. Continuity of $a$ implies the relations
\begin{eqnarray} 
c_+ & = & \frac{e^{H\epsilon}}{\epsilon}~,\\
c_- & = & c_+~ \frac{\tau_{\rm turn} }{ \tilde\tau_{\rm turn}}~.
\end{eqnarray} 

We will assume that events of type $I$ occur only in the terminal vacuum.  We specify that no such events occur at times earlier than $\tau={\delta_+}$, where ${\delta_+} >\epsilon$, or later than $\tilde\tau={\delta_-}$ where ${\delta_-}>0$.  In the intermediate region, we assume that events of type $I$ occur at a constant density $D_I$ per unit four-volume.  For later convenience we choose $D_I=v_*^{-1}$, where $v_*$ is one horizon volume of the de~Sitter vacuum.

The cumulative number of events of type $I$ in regions with expansion factor smaller than $e^\zeta$ generally receives a contribution from both the expanding and contracting region.  It will be convenient to compute these contributions separately:
\begin{equation}
\frac{dN_I}{dN_m}(\zeta)=\frac{dN^+_I}{dN_m}[\tau(\zeta)]
+ \frac{dN^-_I}{dN_m}[\tilde\tau(\zeta)]~.
\end{equation}
The three-volume in the bubble at the time $\tau$ is one parent horizon volume, times the volume expansion factor since nucleation, which is $(c_+\tau)^3$ in the expanding era and $(c_-\tilde\tau)^3$ in the contracting era.  Hence,
\begin{equation}
\frac{dN^+_I}{dN_m}(\tau)=D_I\,v_*\int_{\delta_+}^\tau c_+^3\tau'^3 d\tau 
= \frac{c_+^3}{4} (\tau^4-{\delta_+}^4)
\end{equation}
and
\begin{equation}
\frac{dN^-_I}{dN_m}(\tilde\tau)=D_I\, v_*\int_{{\delta_-}}^\tau c_-^3\tilde\tau'^3 d\tilde\tau 
= \frac{c_-^3}{4} (\tilde\tau^4-{\delta_-}^4)
\end{equation}
Both expressions can be regarded as functions of $\zeta$, by substituting $\tau=e^\zeta/c_+$ and $\tilde\tau=e^\zeta/c_-$ from Eq.~(\ref{eq-zeta}).  The above results hold for ${\delta_+}<\tau<\tau_{\rm turn}$ and ${\delta_-}<\tilde\tau< \tilde\tau_{\rm turn}$, respectively, and in the corresponding range of $\zeta$.  Outside this range, the cumulative number of events receives no contributions and thus remains constant.  Thus, each function is defined in the entire range $-\infty<\zeta<\infty$; for example,
\begin{equation}
\frac{dN^+_I}{dN_m}(\zeta)=\left\{\begin{array}{ll} 
0~, & 0<\zeta<\zeta({\delta_+}) \\
\frac{c_+^3}{4} [\tau(\zeta)^4-{\delta_+}^4]~, & \zeta({\delta_+})<\zeta<\zeta(\tau_{\rm turn}) \\
\frac{c_+^3}{4} [\tau_{\rm turn}^4-{\delta_+}^4]~, & \zeta>\zeta(\tau_{\rm turn})~.
\end{array}\right.
\end{equation}

The probability for events of type $I$ is given by Eq.~(\ref{eq-barnithree}), which reduces to $N_{Im}\kappa_{m*}\bar n_*$ in our model.  Since we consider events that can only occur in the terminal vacuum, the factors $\kappa_{m*}\bar n_*$ will drop out of all relative probabilities,\footnote{Since we have only specified one type of event, $I$, it may seem that there are no relative probabilities to compute.   However, it is trivial to repeat our analysis for some other type of event $J$, for which one might specify different values of ${\delta_+}$ and ${\delta_-}$, or a nonconstant density. The point is that the unnormalized probability is finite; hence, relative probabilities will be well-defined.}  and the unnormalized probability for $I$ is given by
\begin{equation}
N_{Im}= N^+_{Im} + N^-_{Im}~,
\end{equation}
where
\begin{equation}
N^\pm_{Im}=\int_{-\infty}^\infty d\zeta\, e^{-3\zeta}\, \frac{dN^\pm_I}{dN_m}(\zeta)~.
\end{equation}
We have approximated $\gamma=3-\kappa_*\approx 3$.  It is convenient to perform these integrals in terms of $\tau$ and $\tilde\tau$.  One finds
\begin{equation}
N_{Im}^+=\frac{1}{4}\int_{\delta_+}^{\tau_{\rm turn} }\frac{d\tau}{\tau^4}(\tau^4-{\delta_+}^4)+
\frac{1}{4}\int_{\tau_{\rm turn} }^\infty\frac{d\tau}{\tau^4}(\tau_{\rm turn}^4-{\delta_+}^4)=\frac{1}{3}(\tau_{\rm turn} -{\delta_+})
\end{equation}
and 
\begin{equation}
N_{Im}^-=\frac{1}{4}\int_{{\delta_-}}^{\tilde\tau_{\rm turn}} \frac{d\tilde\tau}{\tilde\tau^4}(\tilde\tau^4-{\delta_-}^4)+
\frac{1}{4}\int_{\tilde\tau_{\rm turn} }^\infty\frac{d\tilde\tau}{\tilde\tau^4}(\tilde\tau_{\rm turn}^4-{\delta_-}^4)=
\frac{1}{3}(\tilde\tau_{\rm turn} -{\delta_-})~.
\end{equation}

Note that the latter result is finite, even in the limit as ${\delta_-}\to 0$.  This shows that the total contribution from the infinite number of disconnected future regions converges.  In this example, of course, the caustic occurs only at the spacelike singularity at ${\delta_-=0}$.  Classical spacetime breaks down a finite time before the caustic is reached, so that the limit ${\delta_-}\to 0$ would not strictly be taken in any physical application.  But for caustics in gravitationally bound regions, our result is significant, since it shows that no portions of regular spacetime regions need be excluded to make the New Scale Factor Cutoff finite---not even portions of Planck size, which were excluded in the argument of Sec.~\ref{sec-collapse} for convenience.

The total unnormalized probability for events of type $I$ is proportional to the time duration during which a constant spacetime density for such events is turned on, $\tau_{\rm crunch}-{\delta_+}-{\delta_-}$.  The simplicity of this result suggests that there is a simpler way of deriving it.  In a separate publication~\cite{BouMaiTA}, it will be shown that the New Scale Factor Cutoff is equivalent to the Short Fat Geodesic, a local measure that defines probabilities in terms of the expected number of events of type $I$ in an infinitesimal neighborhood of a single geodesic terminated at the first caustic.  This generalizes the old scale factor time cutoff/fat geodesic duality~\cite{BouFre08b} to arbitrary spacetime regions.

\acknowledgments I am grateful to D.~Mainemer Katz for help with identifying the theorem quoted in Sec.~\ref{sec-def}.  I would also like to thank B.~Freivogel for helpful discussions.  This work was supported by the Berkeley Center for Theoretical Physics, by the National Science Foundation (award numbers 0855653 and 0756174), by fqxi grant RFP3-1004, and by the U.S.\ Department of Energy under Contract DE-AC02-05CH11231.

\bibliographystyle{utcaps}
\bibliography{all}

\providecommand{\href}[2]{#2}\begingroup\raggedright\begin{thebibliography}{10}

\bibitem{Per98}
{\bf Supernova Cosmology Project} Collaboration, S.~Perlmutter {\em et al.},
  ``Measurements of {O}mega and {L}ambda from 42 High-Redshift Supernovae,''
  {\em Astrophys. J.} {\bf 517} (1999)  565--586,
\href{http://arXiv.org/abs/astro-ph/9812133}{{\tt astro-ph/9812133}}.
%%CITATION = ASTRO-PH 9812133;%%.

\bibitem{Rie98}
{\bf Supernova Search Team} Collaboration, A.~G. Riess {\em et al.},
  ``Observational Evidence from Supernovae for an Accelerating Universe and a
  Cosmological Constant,'' {\em Astron. J.} {\bf 116} (1998)  1009--1038,
\href{http://arXiv.org/abs/astro-ph/9805201}{{\tt astro-ph/9805201}}.
%%CITATION = ASTRO-PH 9805201;%%.

\bibitem{GutWei83}
A.~H. Guth and E.~J. Weinberg, ``Could the universe have recovered from a slow
  first-order phase transition?,'' {\em Nucl. Phys.} {\bf B212} (1983)
  321--364.

\bibitem{Car00}
S.~M. Carroll, ``The cosmological constant,''
\href{http://arxiv.org/abs/astro-ph/0004075}{{\tt astro-ph/0004075}}.
%%CITATION = ASTRO-PH 0004075;%%.

\bibitem{Bou12}
R.~Bousso, ``{The Cosmological Constant Problem, Dark Energy, and the Landscape
  of String Theory},''
\href{http://arxiv.org/abs/1203.0307}{{\tt arXiv:1203.0307 [astro-ph.CO]}}.
%%CITATION = ARXIV:1203.0307;%%.

\bibitem{CDL}
S.~Coleman and F.~D. Luccia, ``Gravitational effects on and of vacuum decay,''
  {\em Phys. Rev. D} {\bf 21} (1980)  3305--3315.

\bibitem{GibHaw77a}
G.~W. Gibbons and S.~W. Hawking, ``Cosmological Event Horizons, Thermodynamics,
  and Particle Creation,'' {\em Phys. Rev. D} {\bf 15} (1977)  2738--2751.

\bibitem{BP}
R.~Bousso and J.~Polchinski, ``Quantization of four-form fluxes and dynamical
  neutralization of the cosmological constant,'' {\em JHEP} {\bf 06} (2000)
  006,
\href{http://arxiv.org/abs/hep-th/0004134}{{\tt hep-th/0004134}}.
%%CITATION = JHEPA,0006,006;%%.

\bibitem{FelHal05}
B.~Feldstein, L.~J. Hall, and T.~Watari, ``Density perturbations and the
  cosmological constant from inflationary landscapes,'' {\em Phys. Rev. D} {\bf
  72} (2005)  123506,
\href{http://arxiv.org/abs/hep-th/0506235}{{\tt hep-th/0506235}}.
%%CITATION = HEP-TH 0506235;%%.

\bibitem{BouFre06b}
R.~Bousso and B.~Freivogel, ``A paradox in the global description of the
  multiverse,'' {\em JHEP} {\bf 06} (2007)  018,
\href{http://arxiv.org/abs/hep-th/0610132}{{\tt hep-th/0610132}}.
%%CITATION = HEP-TH/0610132;%%.

\bibitem{Pag06}
D.~N. Page, ``Is our universe likely to decay within 20 billion years?,''
\href{http://arxiv.org/abs/hep-th/0610079}{{\tt hep-th/0610079}}.
%%CITATION = HEP-TH/0610079;%%.

\bibitem{BouFre07}
R.~Bousso, B.~Freivogel, and I.-S. Yang, ``{Boltzmann babies in the proper time
  measure},'' \href{http://dx.doi.org/10.1103/PhysRevD.77.103514}{{\em Phys.
  Rev.} {\bf D77} (2008)  103514},
\href{http://arxiv.org/abs/0712.3324}{{\tt arXiv:0712.3324 [hep-th]}}.
%%CITATION = 0712.3324;%%.

\bibitem{Bou06}
R.~Bousso, ``Holographic probabilities in eternal inflation,'' {\em Phys. Rev.
  Lett.} {\bf 97} (2006)  191302,
\href{http://arxiv.org/abs/hep-th/0605263}{{\tt hep-th/0605263}}.
%%CITATION = HEP-TH/0605263;%%.

\bibitem{Bou09}
R.~Bousso, ``{Complementarity in the Multiverse},''
  \href{http://dx.doi.org/10.1103/PhysRevD.79.123524}{{\em Phys. Rev.} {\bf
  D79} (2009)  123524},
\href{http://arxiv.org/abs/0901.4806}{{\tt arXiv:0901.4806 [hep-th]}}.
%%CITATION = 0901.4806;%%.

\bibitem{BouYan09}
R.~Bousso and I.-S. Yang, ``{Global-Local Duality in Eternal Inflation},''
  \href{http://dx.doi.org/10.1103/PhysRevD.80.124024}{{\em Phys. Rev.} {\bf
  D80} (2009)  124024},
\href{http://arxiv.org/abs/0904.2386}{{\tt arXiv:0904.2386 [hep-th]}}.
%%CITATION = 0904.2386;%%.

\bibitem{LinMez93}
A.~Linde and A.~Mezhlumian, ``Stationary universe,'' {\em Phys. Lett.} {\bf
  B307} (1993)  25--33, \href{http://arxiv.org/abs/gr-qc/9304015}{{\tt
  gr-qc/9304015}}.

\bibitem{LinLin94}
A.~Linde, D.~Linde, and A.~Mezhlumian, ``From the Big Bang theory to the theory
  of a stationary universe,'' {\em Phys. Rev. D} {\bf 49} (1994)  1783--1826,
  \href{http://arxiv.org/abs/gr-qc/9306035}{{\tt gr-qc/9306035}}.

\bibitem{Lin06}
A.~Linde, ``Sinks in the Landscape, {B}oltzmann {B}rains, and the Cosmological
  Constant Problem,'' {\em JCAP} {\bf 0701} (2007)  022,
\href{http://arxiv.org/abs/hep-th/0611043}{{\tt hep-th/0611043}}.
%%CITATION = HEP-TH 0611043;%%.

\bibitem{DesGut08a}
A.~De~Simone, A.~H. Guth, M.~P. Salem, and A.~Vilenkin, ``{Predicting the
  cosmological constant with the scale-factor cutoff measure},''
  \href{http://dx.doi.org/10.1103/PhysRevD.78.063520}{{\em Phys. Rev.} {\bf
  D78} (2008)  063520},
\href{http://arxiv.org/abs/0805.2173}{{\tt arXiv:0805.2173 [hep-th]}}.
%%CITATION = 0805.2173;%%.

\bibitem{FreKle05}
B.~Freivogel, M.~Kleban, M.~Rodriguez~Martinez, and L.~Susskind,
  ``{Observational consequences of a landscape},'' {\em JHEP} {\bf 03} (2006)
  039,
\href{http://arxiv.org/abs/hep-th/0505232}{{\tt arXiv:hep-th/0505232}}.
%%CITATION = HEP-TH/0505232;%%.

\bibitem{Fre08}
B.~Freivogel, ``{Anthropic Explanation of the Dark Matter Abundance},''
\href{http://arxiv.org/abs/0810.0703}{{\tt arXiv:0810.0703 [hep-th]}}.
%%CITATION = 0810.0703;%%.

\bibitem{BouFre08b}
R.~Bousso, B.~Freivogel, and I.-S. Yang, ``{Properties of the scale factor
  measure},''
\href{http://arxiv.org/abs/0808.3770}{{\tt arXiv:0808.3770 [hep-th]}}.
%%CITATION = 0808.3770;%%.

\bibitem{DesGut08b}
A.~De~Simone {\em et al.}, ``{Boltzmann brains and the scale-factor cutoff
  measure of the multiverse},''
  \href{http://dx.doi.org/10.1103/PhysRevD.82.063520}{{\em Phys. Rev.} {\bf
  D82} (2010)  063520},
\href{http://arxiv.org/abs/0808.3778}{{\tt arXiv:0808.3778 [hep-th]}}.
%%CITATION = 0808.3778;%%.

\bibitem{BouHal09}
R.~Bousso, L.~J. Hall, and Y.~Nomura, ``{Multiverse Understanding of
  Cosmological Coincidences},''
  \href{http://dx.doi.org/10.1103/PhysRevD.80.063510}{{\em Phys. Rev.} {\bf
  D80} (2009)  063510},
\href{http://arxiv.org/abs/0902.2263}{{\tt arXiv:0902.2263 [hep-th]}}.
%%CITATION = 0902.2263;%%.

\bibitem{Sal09}
M.~P. Salem, ``{Negative vacuum energy densities and the causal diamond
  measure},''
\href{http://arxiv.org/abs/0902.4485}{{\tt arXiv:0902.4485 [hep-th]}}.
%%CITATION = 0902.4485;%%.

\bibitem{BouFre10e}
R.~Bousso, B.~Freivogel, S.~Leichenauer, and V.~Rosenhaus, ``{Geometric origin
  of coincidences and hierarchies in the landscape},''
  \href{http://dx.doi.org/10.1103/PhysRevD.84.083517}{{\em Phys. Rev.} {\bf
  D84} (2011)  083517},
\href{http://arxiv.org/abs/1012.2869}{{\tt arXiv:1012.2869 [hep-th]}}.
%%CITATION = 1012.2869;%%.

\bibitem{DesSal09}
A.~De~Simone and M.~P. Salem, ``{The distribution of $\Omega_k$ from the
  scale-factor cutoff measure},''
  \href{http://dx.doi.org/10.1103/PhysRevD.81.083527}{{\em Phys.Rev.} {\bf D81}
  (2010)  083527},
\href{http://arxiv.org/abs/0912.3783}{{\tt arXiv:0912.3783 [hep-th]}}.
%%CITATION = ARXIV:0912.3783;%%.

\bibitem{LarNom11}
G.~Larsen, Y.~Nomura, and H.~L.~L. Roberts, ``{The Cosmological Constant in the
  Quantum Multiverse},''
\href{http://arxiv.org/abs/1107.3556}{{\tt arXiv:1107.3556 [hep-th]}}.
%%CITATION = 1107.3556;%%.

\bibitem{Sal12}
M.~P. Salem, ``{The CMB and the measure of the multiverse},''
\href{http://arxiv.org/abs/1204.1569}{{\tt arXiv:1204.1569 [hep-th]}}.
%%CITATION = ARXIV:1204.1569;%%.

\bibitem{Teg05}
M.~Tegmark, ``What does inflation really predict?,'' {\em JCAP} {\bf 0504}
  (2005)  001, \href{http://arxiv.org/abs/astro-ph/0410281}{{\tt
  astro-ph/0410281}}.

\bibitem{SchVil06}
D.~Schwartz-Perlov and A.~Vilenkin, ``Probabilities in the
  {B}ousso-{P}olchinski multiverse,'' {\em JCAP} {\bf 0606} (2006)  010,
\href{http://arxiv.org/abs/hep-th/0601162}{{\tt hep-th/0601162}}.
%%CITATION = HEP-TH 0601162;%%.

\bibitem{Fre11}
B.~Freivogel, ``{Making predictions in the multiverse},''
  \href{http://dx.doi.org/10.1088/0264-9381/28/20/204007}{{\em Class. Quant.
  Grav.} {\bf 28} (2011)  204007},
\href{http://arxiv.org/abs/1105.0244}{{\tt arXiv:1105.0244 [hep-th]}}.
%%CITATION = ARXIV:1105.0244;%%.

\bibitem{Sal11}
M.~P. Salem, ``{Bubble collisions and measures of the multiverse},'' {\em JCAP}
  {\bf 1201} (2012)  021,
\href{http://arxiv.org/abs/1108.0040}{{\tt arXiv:1108.0040 [hep-th]}}.
%%CITATION = ARXIV:1108.0040;%%.

\bibitem{BouFre10b}
R.~Bousso, B.~Freivogel, S.~Leichenauer, and V.~Rosenhaus, ``{Boundary
  definition of a multiverse measure},''
\href{http://arxiv.org/abs/1005.2783}{{\tt arXiv:1005.2783 [hep-th]}}.
%%CITATION = 1005.2783;%%.

\bibitem{BouHar07}
R.~Bousso, R.~Harnik, G.~D. Kribs, and G.~Perez, ``Predicting the cosmological
  constant from the causal entropic principle,'' {\em Phys. Rev. D} {\bf 76}
  (2007)  043513,
\href{http://arxiv.org/abs/hep-th/0702115}{{\tt hep-th/0702115}}.
%%CITATION = HEP-TH 0702115;%%.

\bibitem{BouHar10}
R.~Bousso and R.~Harnik, ``{The Entropic Landscape},''
  \href{http://dx.doi.org/10.1103/PhysRevD.82.123523}{{\em Phys. Rev.} {\bf
  D82} (2010)  123523},
\href{http://arxiv.org/abs/1001.1155}{{\tt arXiv:1001.1155 [hep-th]}}.
%%CITATION = 1001.1155;%%.

\bibitem{BouFre10d}
R.~Bousso, B.~Freivogel, S.~Leichenauer, and V.~Rosenhaus, ``{A geometric
  solution to the coincidence problem, and the size of the landscape as the
  origin of hierarchy},''
  \href{http://dx.doi.org/10.1103/PhysRevLett.106.101301}{{\em Phys. Rev.
  Lett.} {\bf 106} (2011)  101301},
\href{http://arxiv.org/abs/1011.0714}{{\tt arXiv:1011.0714 [hep-th]}}.
%%CITATION = 1011.0714;%%.

\bibitem{Wei87}
S.~Weinberg, ``Anthropic Bound on the Cosmological Constant,''
{\em Phys. Rev. Lett.} {\bf 59} (1987)  2607.
%%CITATION = PRLTA,59,2607;%%.

\bibitem{MarSha97}
H.~Martel, P.~R. Shapiro, and S.~Weinberg, ``Likely Values of the Cosmological
  Constant,''
\href{http://arxiv.org/abs/astro-ph/9701099}{{\tt astro-ph/9701099}}.
%%CITATION = ASTRO-PH 9701099;%%.

\bibitem{Wald}
R.~M. Wald, {\em General Relativity}.
\newblock The University of Chicago Press, Chicago, 1984.

\bibitem{HawEll}
S.~W. Hawking and G.~F.~R. Ellis, {\em The large scale stucture of space-time}.
\newblock Cambridge University Press, Cambridge, England, 1973.

\bibitem{BouSus11}
R.~Bousso and L.~Susskind, ``{The Multiverse Interpretation of Quantum
  Mechanics},''
\href{http://arxiv.org/abs/1105.3796}{{\tt arXiv:1105.3796 [hep-th]}}.
%%CITATION = 1105.3796;%%.

\bibitem{CEB1}
R.~Bousso, ``A covariant entropy conjecture,'' {\em JHEP} {\bf 07} (1999)  004,
\href{http://arxiv.org/abs/hep-th/9905177}{{\tt hep-th/9905177}}.
%%CITATION = JHEPA,9907,004;%%.

\bibitem{BouFre10a}
R.~Bousso, B.~Freivogel, and S.~Leichenauer, ``{Saturating the holographic
  entropy bound},'' \href{http://dx.doi.org/10.1103/PhysRevD.82.084024}{{\em
  Phys. Rev.} {\bf D82} (2010)  084024},
\href{http://arxiv.org/abs/1003.3012}{{\tt arXiv:1003.3012 [hep-th]}}.
%%CITATION = 1003.3012;%%.

\bibitem{GarSch05}
J.~Garriga, D.~Schwartz-Perlov, A.~Vilenkin, and S.~Winitzki, ``Probabilities
  in the inflationary multiverse,'' {\em JCAP} {\bf 0601} (2006)  017,
\href{http://arxiv.org/abs/hep-th/0509184}{{\tt hep-th/0509184}}.
%%CITATION = HEP-TH 0509184;%%.

\bibitem{BouMaiTA}
R.~Bousso and D.~Mainemer {\em (to appear)}  .

\end{thebibliography}\endgroup


\providecommand{\href}[2]{#2}\begingroup\raggedright\endgroup

\end{document}